\documentclass[aps, prl, twocolumn,showkeys, showpacs,amsmath,amssymb,superscriptaddress]{revtex4}

\usepackage{graphicx}
\usepackage{dcolumn}
\usepackage{bm}
\usepackage{epsfig}
\usepackage{color}

\usepackage{amsmath}

\begin{document}
\title{Non-Perturbative  Hydrodynamic Limits}


\author{I. V. Karlin} \email{karlin@lav.mavt.ethz.ch}
\affiliation{Aerothermochemistry and Combustion Systems Lab, ETH Z\"urich, 8092 Z\"urich, Switzerland}
\author{S. S. Chikatamarla} \email{chikatamarla@lav.mavt.ethz.ch}
\affiliation{Aerothermochemistry and Combustion Systems Lab, ETH Z\"urich, 8092 Z\"urich, Switzerland}
\author{M. Kooshkbaghi} \email{kooshkbaghi@lav.mavt.ethz.ch}
\affiliation{Aerothermochemistry and Combustion Systems Lab, ETH Z\"urich, 8092 Z\"urich, Switzerland}

%

\date{\today}
\begin{abstract}
We introduce non-perturbative analytical techniques for the derivation of the hydrodynamic manifolds from kinetic equations.
The new approach is analogous to the Schwinger-Dyson equation of quantum field theories,
and its derivation is demonstrated with the construction of the exact diffusion manifold for a model kinetic equation.
The novel approach is superior to the classical Chapman-Enskog method.
\end{abstract}


\maketitle

Derivation of hydrodynamic equations from the Boltzmann kinetic equation is the classical problem of statistical mechanics.
The best known techniques, the Chapman-Enskog (CE) method \cite{ChapK}, is a perturbation method based on a small parameter (Knudsen number).
While the formal derivation of the Navier-Stokes equation from the Boltzmann equation by this method is a textbook example of the success of statistical physics, many problems related to the hydrodynamic limit of the kinetic equations remained unsolved \cite{StRaymond2009}. The CE perturbation expansion fails as the post-Navier-Stokes hydrodynamic equations (Burnett's equations) are unstable \cite{Bobylev1982}. Moreover, even at the Navier-Stokes level, the rate of dissipation (by viscosity, thermal conductivity or diffusion) is unbounded which contradicts the finiteness of relaxation times in the kinetic picture. Finally, as was pointed out by many authors \cite{Slemrod2012}, the smallness parameter of the problem is not a fixed quantity (unlike, for example, the fine structure constant in quantum electrodynamics) but can be always scaled out. 

All this points at the inability of the CE method to tackle the above problem, and non-perturbative approaches are sought.
To that end, minimal kinetic theories such as Grad's finite-moment systems have been studied in details \cite{Karlin2002}, including {\it exact} summation of the entire CE series. Some results obtained along these lines are surprising. For example, Slemrod \cite{Slemrod2012} noted that exact summation of the CE expansion results in hydrodynamic equations of Korteweg's type (that is, containing the capillarity-type contribution) rather than a (modified) Navier-Stokes equation. However, results for kinetic equations, that is, for infinite moment systems remained almost entirely unexplored due to lack of  analytical non-perturbative techniques.

In this paper, we introduce a novel analytic approach to extracting non-perturbatively the hydrodynamic component out of kinetic equations.
We consider in detail a model kinetic equation, and derive exact invariance equation which is in striking resemblance with the Schwinger-Dyson equation of non-perturbative field theories.
Based on the exact solution to the invariance equation, we build up a systematic non-perturbative extension procedure and show its relation to the conventional Chapman-Enskog method.
Present approach can be used beyond the model kinetic equation considered below, and we outline some further steps towards the non-perturbative hydrodynamic limit.

\begin{figure}[!tbp]
\begin{center}
\includegraphics[width=0.45\textwidth,clip]{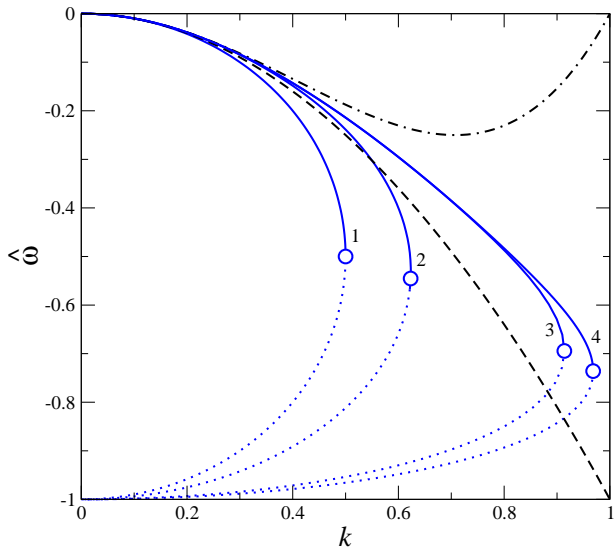}
\end{center}
  \caption{(Color online) Hydrodynamic limit of Eq.\ (\ref{eq:BGK_diffusion}). Dash: First CE approximation, $\hat{\omega}_{\rm CE}^{(2)}=-k^2$ (unbounded as $k\to\infty$); Dot-dash: Burnett-type approximation, $\hat{\omega}_{\rm CE}^{(4)}=-k^2+k^4$ (unstable at $k>1$); Line: Continuation by the sequence (\ref{eq:pullout}). Curves 1, 2, 3 and 4 correspond to the hydrodynamic branch
	$\hat{\omega}_{\rm H}^{(2n)}$ for $n=1,2,20,25$, respectively.
	 Interception by a partner kinetic mode $\hat{\omega}_{\rm P}^{(2n)}$ (dots) at $k=k_{\rm c}^{(2n)}$ is indicated by open circle. }
\label{fig:1}
\end{figure}

We consider the one-dimensional kinetic equation for the distribution function $f(x,v,t)$:
\begin{equation}\label{eq:BGK_diffusion}
\partial_t f=-v\partial_xf-{\tau}^{-1}\left(f-f^{\rm eq}\right),
\end{equation}
where the local equilibrium has the form
\begin{equation}
f^{\rm eq}=n(x,t)\left(2\pi k_BT/m\right)^{-1/2}e^{-\frac{mv^2}{2k_BT}},
\end{equation}
with $n(x,t)$ the locally conserved density,
\begin{equation}
n=\int_{-\infty}^{\infty} f(v,x,t)dv.
\end{equation}
The conventional CE analysis of (\ref{eq:BGK_diffusion}) is to introduce a small parameter $\epsilon$: $\tau\to\varepsilon\tau$, and to expand, $f=f^{(0)}+\varepsilon f^{(1)}+\dots$, in order to produce a closure to the balance equation, $\partial_t n=-\partial_x j$, where $j=\int_{-\infty}^{\infty}vfdv$ is the density flux. One easily computes the first two terms of the CE closure, $\partial_tn=\omega_{\rm CE}n$: $\omega_{\rm CE}^{(2)}=\varepsilon D\partial_x^2$ (first approximation; this is the conventional diffusion); $\omega_{\rm CE}^{(4)}=\varepsilon D\partial_x^2+\varepsilon^2 D^2\partial_x^4$ (second approximation), where $D=\tau k_BT/m$ is the diffusion coefficient. The problem with such such a perturbation approach is readily seen even in the present case: After an appropriate rescaling in time and space, the mode's relaxation rate $\hat{\omega}(k)$, where $k$ is the wave vector, is (a) Unbounded as $k\to \infty$ in the first approximation, and (b) Unstable after $k>1$ in the second approximation (see Fig.\ \ref{fig:1}).
Thus, even for the simplest kinetic equation (\ref{eq:BGK_diffusion}), application of the perturbative CE expansion inherits the essential problems as in the case of the Boltzmann equation. Our goal here is to develop a non-perturbative method to study the hydrodynamic limit of (\ref{eq:BGK_diffusion}).

In the sequel we put $k_BT/m=1$, $\tau=1$.
Moments of the distribution functions can be written,
\begin{equation}\label{eq:moments}
M_{l}(x,t)=\int_{-\infty}^{\infty} v^l f({v}, x,t)d{v}.
\end{equation}
Equation (\ref{eq:BGK_diffusion}) is equivalent to an infinite moment system,
\begin{equation}\label{eq:chain}
\partial_t M_{l}=-\partial_x M_{l+1}-M_{l}+M_{l}^{\rm eq}.
\end{equation}
Instead of the infinite set of moments  (\ref{eq:moments}), it proves convenient to consider the generating function $Z({\lambda},{x},t)$ (Fourier transform in the velocity variable),
\begin{equation}
Z=\int_{-\infty}^{\infty} e^{-i{\lambda}{v}}f({v},{x},t)d{v}.
\end{equation}
Let us denote $Z^{\pm}$ the real and imaginary parts of
$Z=Z^++iZ^-$;
Since moments (\ref{eq:moments}) are real-valued, for even and odd moments we have, respectively,
\begin{equation}\nonumber\begin{split}
&M_{2n}=(-1)^{n}\left(\partial^{2n}_\lambda Z^+\right)_{\lambda=0},\\
&M_{2n+1}=(-1)^{n+1}\left(\partial^{2n+1}_\lambda Z^-\right)_{\lambda=0}.\end{split}
\end{equation}
In terms of the generating function, Eq.\ (\ref{eq:BGK_diffusion}) is represented as a coupled system for $Z^+$ and $Z^-$,
\begin{equation}
\begin{split}\label{eq:eq4generator}
&\partial_t Z^+  =  \partial_x\partial_\lambda Z^- -Z^+ +Z^{\rm eq},\\
&\partial_t Z^-  = -\partial_x\partial_\lambda Z^+ -Z^-,
\end{split}
\end{equation}
where
$Z^{\rm eq}=n(x,t)\varphi^{\rm eq}$, and $\varphi^{\rm eq}=e^{-\frac{\lambda^2}{2}}$.
Finally, applying Fourier transform in space, equations (\ref{eq:eq4generator}) become
\begin{eqnarray}
\partial_t \hat{Z}^+  &=& {ik}\partial_\lambda \hat{Z}^- -\hat{Z}^+ +\hat{n}\varphi^{\rm eq},\label{eq:Zp}\\
\partial_t \hat{Z}^-  &=& -{ik}\partial_\lambda \hat{Z}^+ -\hat{Z}^-,\label{eq:Zm}
\end{eqnarray}
where $\hat{Z}^{\pm}(\lambda,k,t)=\int_{-\infty}^{\infty} e^{-ikx}Z^{\pm}(\lambda,x,t)dx$ depend on the wave vector $k$.

Equations (\ref{eq:Zp},\ref{eq:Zm}) are the starting point for the non-perturbative analysis.
In the hydrodynamic limit, all moments depend on space and time only through their dependence on the locally conserved field (density).
To this end, the most general and yet unknown relation for the generating functions $\hat{Z}^{\pm}$ can be written,
\begin{eqnarray}
\hat{Z}^+&=&\hat{\Theta}^+(\lambda, k^2)\hat{n}(k,t),\\
\hat{Z}^-&=&ik\hat{\Theta}^-(\lambda, k^2)\hat{n}(k,t),
\end{eqnarray}
where $\hat{\Theta}^{\pm}$ are functions in question, satisfying the consistency conditions,
$\hat{\Theta}^+(0,k^2)=1$, $\hat{\Theta}^-(0,k^2)=0$.
Knowing $\hat{\Theta}^-$, the balance equation becomes,
$\partial_t\hat{n}=-k^2\hat{G}\hat{n}$,
where
\begin{equation}
\hat{G}=(\partial_\lambda \hat{\Theta}^-)_{\lambda=0},
\end{equation}
is the extended diffusion coefficient (EDC).

\begin{figure}[!tbp]
\begin{center}
\includegraphics[width=0.45\textwidth,clip]{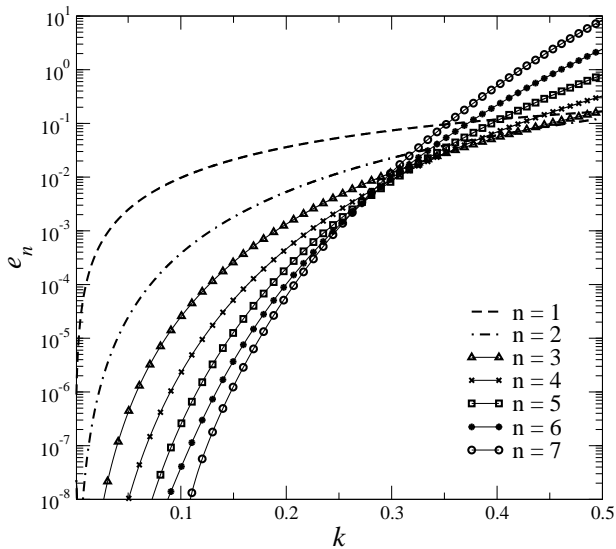}
\end{center}
  \caption{Deviation of the CE approximations 
	$\hat{\omega}_{\rm CE}^{(2n)}$ 
	         for $n=1,\dots,7$ from the exact solution, $e_n=\left|({\hat{\omega}^{(50)}_{\rm H}-\hat{\omega}_{\rm CE}^{(2n)}})/{\hat{\omega}^{(50)}_{\rm H}}\right|$.}
\label{fig:2}
\end{figure}

Let us now formulate the most general condition for  $\hat{\Theta}^{\pm}$. The time-derivative of the generating function can be computed in two different ways.
On the one hand, it is computed by chain rule and using the balance equation:
\begin{eqnarray}
\partial_t^{\rm macro} \hat{Z}^+&=&\frac{\partial \hat{Z}^+}{\partial \hat{n}}\partial_t \hat{n}
                                =(-{k^2}\hat{G}\hat{\Theta}^+)\hat{n},\\
\partial_t^{\rm macro} \hat{Z}^-&=&\frac{\partial \hat{Z}^-}{\partial \hat{n}}\partial_t \hat{n}
                                =ik(-{k^2}\hat{G}\hat{\Theta}^-)\hat{n}.
\end{eqnarray}
This is the macroscopic time derivative, or the derivative of the yet unknown closed generating function due to the (also yet unknown) closed balance equation. On the other hand, the microscopic time derivative is given by the right hand side of (\ref{eq:Zp},\ref{eq:Zm}):
\begin{eqnarray}
\partial_t^{\rm micro} \hat{Z}^+&=&(-{k^2}\partial_\lambda \hat{\Theta}^- -\hat{\Theta}^+ +\varphi^{\rm eq})\hat{n},\\
\partial_t^{\rm micro} \hat{Z}^-&=&ik(-\partial_\lambda \hat{\Theta}^+ -\hat{\Theta}^-)\hat{n}.
\end{eqnarray}
The dynamic invariance condition \cite{Karlin2002} requires that the micro- and the macroscopic derivatives of the generating function should give the same result, independently of $\hat{n}$:
\begin{equation}
\partial_t^{\rm micro} \hat{Z}^{\pm}=\partial_t^{\rm macro} \hat{Z}^{\pm}.
\end{equation}
Thus, the invariance condition for the generating function is  a system of two first-order equations,
\begin{eqnarray}
&&-k^2\hat{G}\hat{\Theta}^+ +k^2\partial_\lambda \hat{\Theta}^- +\hat{\Theta}^+-\varphi^{\rm eq}=0,\label{eq:1}\\
&&-k^2\hat{G}\hat{\Theta}^- +\partial_\lambda \hat{\Theta}^+ +\hat{\Theta}^-=0,\label{eq:2}
\end{eqnarray}
subject to initial conditions, $\hat{\Theta}^+(0,k^2)=1$, $\hat{\Theta}^-(0,k^2)=0$.
It can be readily checked that the invariance equation generates the CE solution when functions $\hat{\Theta}^{\pm}$ are expanded into Taylor series around $k^2=0$.
Our goal is, however, to solve the invariance equation avoiding any expansion of this kind.
We proceed with a few transformations:
(i) Differentiate (\ref{eq:1}) with respect to $\lambda$ and eliminate $\partial_\lambda \hat{\Theta}^+$ to get second-order equation for $\hat{\Theta}^-$:
\begin{equation}\label{eq:Theta_m_2order}
k^2\partial^2_\lambda\hat{\Theta}^- -(1-k^2\hat{G})^2\hat{\Theta}^--\partial_\lambda \varphi^{\rm eq}=0;
\end{equation}
(ii) Differentiate  equation (\ref{eq:Theta_m_2order}) one more time;
Use transformed variable,
$\partial_\lambda \hat{\Theta}^-=\hat{\Sigma}e^{-\lambda^2/2}$, and
note that
$\hat{G}=(\hat{\Sigma})_{\lambda=0}$.
Thus, the invariance equation for the generating function becomes,
\begin{equation}\label{eq:result1}
(k^2\hat{G}-1)^2\hat{\Sigma} +  (1\!-\!\lambda^2)(k^2\hat{\Sigma}-1) \!=\! k^2\!(\partial^2_\lambda \hat{\Sigma}
-2\lambda\partial_\lambda \hat{\Sigma}).
\end{equation}
Invariance equation (\ref{eq:result1}) is key.
We note that, although not form-identical, equation (\ref{eq:result1}) can be regarded an analog of the basic equation of the non-perturbative approach in quantum field theories, the Schwinger-Dyson equation (SDE) \cite{Dyson1949,Schwinger1951}.
Indeed, SDE is a relationship between the one-particle Green's function $\hat{G}$ and the mass operator $\hat{\Sigma}$. Standard derivation of SDE proceeds along the lines similar to the above, considering the generating function of many-particle Green's functions.
In the present context, the ``mass operator" $\hat{\Sigma}$ in (\ref{eq:result1}) provides the coupling to all higher-order moments, and self-consistently defines the EDC $\hat{G}$.
For the present analysis of (\ref{eq:result1}), it is convenient to introduce frequency function
$\hat{\Omega}=-k^2 \hat{\Sigma}$, so that $\hat{\omega}=-k^2\hat{G}=(\hat{\Omega})_{\lambda=0}$:
%
\begin{equation}\label{eq:result1a}
(\hat{\omega} +1)^2\hat{\Omega} \!+\!  k^2(1-\lambda^2)(\hat{\Omega} +1)\! =\! k^2\!(\partial^2_\lambda \hat{\Omega}
\!-\!2\lambda\partial_\lambda \hat{\Omega}).
\end{equation}
Solution to the ODE (\ref{eq:result1a}) with the initial conditions, $(\hat{\Omega})_{\lambda=0}=\hat{\omega}$,
$(\partial_{\lambda}\hat{\Omega})_{\lambda=0}=0$, is found in closed form,
\begin{equation}\begin{split}\label{eq:manifold}
\hat{\Omega}=(\hat{\omega}+1)e^{\frac{\lambda^2}{2}}\cosh(\lambda\beta) -1
 + \frac{\sqrt{2\pi}}{4}\beta e^{\frac{\lambda^2+\beta^2}{2}}\times\\
\!\left[2\cosh(\!\lambda\beta\!){\rm erf}\!\!\left(\!\frac{\beta}{\sqrt{2}}\!\right) \!\!+\!e^{\!-\lambda\beta}\!{\rm erf}\!\!\left(\!\!\frac{\lambda-\beta}{\sqrt{2}}\!\right)\!-\!e^{\lambda\beta}\!{\rm erf}\!\!\left(\!\!\frac{\lambda+\beta}{\sqrt{2}}\!\right)\!\right]
\end{split}
\end{equation}
where $\beta=\sqrt{(\hat{\omega}+1)^2/k^2}$, and ${\rm erf}$ is error function.
Function $\hat{\Omega}(\lambda,\hat{\omega},k^2)$ (\ref{eq:manifold}) describes all invariant manifolds of the kinetic equation: for every fixed $k$, it is a parametric set of functions (of $\lambda$) parameterized by frequency $\hat{\omega}$. The {\it hydrodynamic manifold} is generated by a specific dependence $\hat{\omega}_{\rm H}(k)$ which continues the corresponding solution at $k=0$ to $k>0$. We note that the condition $(\hat{\Omega})_{\lambda=0}=\hat{\omega}$ results in the identity, $\hat{\omega}=\hat{\omega}$, rather than in an equation for $\hat{\omega}_{\rm H}$. Therefore, a special procedure is needed for deriving the function $\hat{\omega}_{\rm H}$.
As we shall see it below, the analyticity of $\hat{\Omega}$ implies that the hydrodynamic manifold indeed extends to $k>0$.
It is instructive to write the solution (\ref{eq:manifold}) in terms of a series,
\begin{equation}\label{eq:manifold_series}
\hat{\Omega}=\hat{\omega}+\sum_{n=1}^{\infty}\frac{\lambda^{2n}\hat{\omega}_{2n}(\hat{\omega},k^2)}{(2n)!(k^{2n})},
\end{equation}
where the coefficients $\hat{\omega}_{2n}$ have the following form:
\begin{equation}\label{eq:coeff_general}
\hat{\omega}_{2n}=(\hat{\omega}+1)[\hat{\omega}(\hat{\omega}+1)^{2n-1}+(2n-1)!!(k^{2n})+\hat{p}_{2n}].
\end{equation}
Here $\hat{p}_{2n}(\hat{\omega},k^2)$ is a polynomial in $\hat{\omega}$ and $k^2$ of the order $k^{2(n-1)}$,
and
$(2n-1)!!=1\cdot 3\cdot 5\dots (2n-1)$.
First few coefficients have the following explicit form:
\begin{eqnarray}
&&\hat{\omega}_2=(\hat{\omega}+1)[\hat{\omega}(\hat{\omega}+1)+k^2],\nonumber\\
&&\hat{\omega}_4=(\hat{\omega}+1)[\hat{\omega}(\hat{\omega}+1)^3+3k^4 +k^2(\hat{\omega}+1)(1+6\hat{\omega})].\nonumber
\end{eqnarray}
We note in passing that coefficients $\hat{\omega}_{2n}$ (\ref{eq:coeff_general}) can be derived directly from the invariance equation (\ref{eq:result1a}), without solving it explicitly. 
Accordingly, the continuation procedure described below can be used also in other cases where analytic solutions to the corresponding invariance equations are difficult to obtain.
Coefficients (\ref{eq:coeff_general}) imply the two limits:
\begin{eqnarray}
&&\hat{\omega}_{2n}\to \hat{\omega}(\hat{\omega}+1)^{2n}, \ k\to 0,\label{eq:lim_0}\\
&&\hat{\omega}_{2n}\to (\hat{\omega}+1)(2n-1)!!(k^{2n}),\ k\to \infty.\label{eq:lim_infty}
\end{eqnarray}
Consequently, we have, at $k\to 0$:
\begin{equation}
\hat{\Omega}_{0}\sim\hat{\omega}\cosh[\lambda (\hat{\omega}+1)/k].
\end{equation}
Requirement of finiteness of the above expression at $k=0$ selects two values for $\omega$: $\hat{\omega}_{\rm H}=0$, and $\hat{\omega}_{\rm K}=-1$. The former is the seed of the hydrodynamic branch, while the latter is the (infinitely degenerated) eigenvalue of the relaxation term of (\ref{eq:BGK_diffusion}).
On the other hand, in the opposite limit $k\to\infty$ (\ref{eq:lim_infty}), the function $\hat{\Omega}$ remains analytic, 
\begin{equation}
\hat{\Omega}_{\infty}=\lim_{k\to\infty}\hat{\Omega}=(1+\hat{\omega})e^{\lambda^2/2}-1.
\end{equation}
Now, since at any $k\ne0$ function $(\ref{eq:manifold})$ is analytic function of $\lambda$, the series (\ref{eq:manifold}) is convergent; hence,
$\hat{\omega}_{2n}(\hat{\omega},k^2)/((2n)!k^{2n})\to 0$ as $n\to\infty$.
This observation implies the following  practical recipe for the continuation for { finite} $k$: Let us consider a sequence of algebraic equations,
\begin{equation}\label{eq:pullout}
\hat{\omega}_{2n}(\hat{\omega},k^2)=0, \ n=1,\ 2,\ \dots
\end{equation}
At $k=0$, for every $n$, equation (\ref{eq:pullout}) seeds one (hydrodynamic) branch at $\hat{\omega}_{\rm H}=0$ and $2n$ degenerated kinetic branches at $\hat{\omega}_{\rm K}=-1$. The solution $\hat{\omega}_{\rm H}^{(2n)}(k)$ with the asymptotics $\hat{\omega}_{\rm H}^{(2n)}(0)=0$ is the extension of the hydrodynamic branch at the $n$th order of the said procedure. In other words, instead of the CE expansion in terms of $k^2$ we consider a sequence of finite-dimensional algebraic problems of increasing order (\ref{eq:pullout}) based on the convergence of the series (\ref{eq:manifold_series}).
We term this a {\it pullout} procedure for the reason clarified in Fig.\ \ref{fig:1}: At each step $n$, the hydrodynamic branch $\hat{\omega}^{(2n)}_{\rm H}$ is pulled out till the critical value $k_{\rm c}^{(2n)}$. At $k_{\rm c}^{(2n)}$, the hydrodynamic branch is intercepted by one of the kinetic branches (partner kinetic mode $\hat{\omega}_{\rm P}^{(2n)}$ with the asymptotics,  $\hat{\omega}_{\rm P}^{(2n)}\to -1$ as $k\to 0$). After the interception, the pair of real-valued solutions $\{\hat{\omega}_{\rm H}^{(2n)},\hat{\omega}_{\rm P}^{(2n)}\}$ continue as the pair of complex-conjugated roots of (\ref{eq:pullout}).
This effect of interception is well known from previously exact summations of the CE expansion of the diffusion-type modes for finite-moment systems \cite{Karlin2002}.
The pullout procedure thus furnishes a non-perturbative extension of the hydrodynamics with monotonically increasing accuracy for finite $k$. At any step of the procedure (\ref{eq:pullout}) the result is bounded, and the interception point $k_{\rm c}^{(2n)}$ increases monotonically. 
In Tab.\ \ref{tab:1}, the matching of the polynomial expansion to order $k^{14}$ for the sequence of pullouts $\hat{\omega}_{\rm H}^{(2n)}$ is verified against the CE expansion. However, the present procedure demonstrates much better and controlled convergence. It is evident from Fig.\ \ref{fig:1} that the hydrodynamic branch is pulled out smoothly so that the result of the higherst order of approximation shown in Fig.\ \ref{fig:1} can be regarded exact up to $k\approx 0.8$. Using the data of Fig.\ \ref{fig:1},  in Fig.\ \ref{fig:2} we present the deviation of the CE expansion at various orders of approximation in therms of $k^2$. While the CE expansion indeed systematically improves the accuracy at very small $k$, this comes at a price of increased deviation at larger $k$. 
By the contrast, the pullout procedure is, in fact, a non-perturbative (that is, a non-polynomial in $k$) method which approximates the infinite-dimensional problem (the infinite moment system) with a sequence of finite-dimensional problems (\ref{eq:pullout}) of increasing order; at each step the available piece of the hydrodynamic manifold is bounded and well controlled. 


We conclude this paper with a discussion.
Above, we have considered the kinetic equation (\ref{eq:BGK_diffusion}) the conventional hydrodynamic limit of which is considered to be the diffusion equation, while its standard perturbative anylysis inherits the main problems. 
We have derived invariance equation, as the analog of the Schwinger-Dyson equation in the present context. 
Exact solution to the invariance equation found above is, to the best of our knowledge, the first of its kind for infinite-dimensional kinetic systems.
Based on the analyticity of the exact solution we have introduced a new procedure of continuation of the hydrodynamic mode and have shown its consistency with the standard Chapman-Enskog expansion.
The change of the perspective on the derivation of hydrodynamics, by switching from a perturbative gradient expansion (expansion in terms of wave vector $k$) to a sequence of finite-order, non-perturbative in gradients problems keeps the extension of the hydrodynamics under control. 
Findings of this work lay out the way to anylyse the reduced description for other systems. In particular, 
related linearized kinetic models such as the Bhatnagar-Gross-Krook kinetic equations can be considered straightforwardly along the above lines, also in three dimensions since the general tensorial structure of the generating function is known in that case \cite{Karlin2002}. 
For the linearized Boltzmann equation, the resulting invariance equation contains then the linearized Boltzmann operator, and the pullout procedure will amount to solving linear integral equations of familiar type \cite{ChapK}. 
Finally, for the non-linear case,  approximate hydrodynamic manifolds arizing in the pullout procedure can be obtained along the lines of non-linear finite-moment systems \cite{Karlin2002}. 
IVK gratefully acknowledges support by 
ERC Advanced Grant 291094-ELBM; MK was supported by SNSF Project 137771.


\begin{table}[!b]
\begin{tabular}{l|l|l|l|l|l|l|l}
& $a_2$ & $a_4$ & $a_6$ & $a_8$ & $a_{10}$ & $a_{12}$ &
$a_{14}$\tabularnewline
\hline
$\hat{\omega}^{(2)}_{\rm H}$ & \boxed{-1} & -1 & -2 & -5 & -14 & -42 & -132\tabularnewline
$\hat{\omega}^{(4)}_{\rm H}$ & \boxed{-1} & \boxed{1} & \boxed{-4 }& 3 & 16 & -122 & 312\tabularnewline
$\hat{\omega}^{(6)}_{\rm H}$ & \boxed{-1} & \boxed{1} & \boxed{-4 }& \boxed{27} & \boxed{-248} & 2110 & -17352\tabularnewline
$\hat{\omega}^{(8)}_{\rm H}$ & \boxed{-1} & \boxed{1} & \boxed{-4 }& \boxed{27} & \boxed{-248} & \boxed{2830} & \boxed{-38232}\tabularnewline
%
$\hat{\omega}_{\rm CE}$ & {-1} & {1} & {-4 }& {27} & {-248} & {2830} & {-38232}\tabularnewline
\end{tabular}
\caption{Expansion of the hydrodynamic mode $\hat{\omega}_{\rm H}=\sum_{n=1}^{\infty}a_{2n}k^{2n}$ by the sequence (\ref{eq:pullout}). Coefficients in boxes match the CE expansion $\hat{\omega}_{\rm CE}$.}\label{tab:1}
\end{table}


\end{document}